\documentclass[final,5p,times,twocolumn,UTF8]{elsarticle}

\usepackage{amssymb}
\usepackage{amsthm}
\usepackage{amsmath}
\usepackage{wallpaper}
\usepackage{dcolumn}
\usepackage{bm}
\usepackage{pdfpages}
\usepackage{graphics}
\usepackage[colorlinks,plainpages=false]{hyperref}
\usepackage{bbding}
\usepackage{booktabs}
\usepackage{lscape}
\usepackage{multirow}
\usepackage{balance}

\journal{Physics Letters B}

\begin{document}

\begin{frontmatter}

\title{Conservation and breaking of pseudospin symmetry}

\author[SunAdr1,SunAdr2]{Ting-Ting Sun\corref{cor1}}
\ead{ttsunphy@zzu.edu.cn}

\author[LiAdr1]{Zhi Pan Li \corref{cor1}}
\ead{zpliphy@swu.edu.cn}

\author[RingAdr1]{Peter Ring \corref{cor1}}
\ead{peter.ring@tum.de}

\cortext[cor1]{Corresponding author.}
\address[SunAdr1]{School of Physics and Microelectronics, Zhengzhou University, Zhengzhou 450001, China}
\address[SunAdr2]{Guangxi Key Laboratory of Nuclear Physics and Nuclear Technology, Guangxi Normal University, Guilin 541004,
China}
\address[LiAdr1]{School of Physical Science and Technology, Southwest University, Chongqing 400715, China}
\address[RingAdr1]{Physik-Department der Technischen Universit\"{a}t M\"{u}nchen, D-85748 Garching, Germany}

\begin{abstract}
Pseudospin symmetry~(PSS)~is a relativistic dynamical symmetry connected
with the lower component of the Dirac spinor. Here, we investigate the
conservation and breaking of PSS in the single-nucleon resonant states, as
an example, using Green's function method that provides a novel way to
precisely describe not only the resonant energies and widths but also the
spacial density distributions for both narrow and wide resonances. The PSS
restoration and breaking are perfectly displayed in the evolution of
resonant parameters and density distributions with the potential depth: In
the PSS limit, i.e., when the attractive scalar and repulsive vector
potentials have the same magnitude but opposite sign, PSS is exactly
conserved with strictly the same energy and width between the PS partners as
well as identical density distributions of the lower components. As the
potential depth increases, the PSS is broken gradually with energy and width
splittings and a phase shift in the density distributions.
\end{abstract}


\end{frontmatter}

\section{\label{sec:intro}Introduction}

Symmetries in the single-particle spectrum of atomic nuclei are of great
importance on nuclear structures and have been extensively studied in the
literature (see Refs.~\cite{PhysRep2005Ginocchio,PhysRep2015HZLiang} and
references therein). More than 50 years ago, pseudospin symmetry (PSS) was
found in atomic nuclei, i.e., the two single-particle states with quantum
numbers ($n,l,j=l+1/2$) and ($n-1,l+2,j=l+3/2$) are quasi-degeneracy and can
be redefined as the pseudospin~(PS) doublets ($\tilde{n}=n$, $\tilde{l}=l+1$,
$j=\tilde{l}\pm 1/2$)~\cite{NPA1969Hecht_137_129,PLB1969Arima_30_517}. The
pseudospin symmetry (PSS) has been used to explain a number of phenomena in
nuclear structures, such as deformation~\cite{PhysScr1982Bohr_26_267},
superdeformation~\cite{PRL1987Dudek_59_1405,PRL1992Bahri_68_2133}, identical
rotational bands~\cite{PRL1990Nazarewicz_64_1654, PRL1990Byrski_64_1650},
magnetic moment~\cite{PRC1999Ginocchio}, quantized
alignment~\cite{PRL1990Stephens} and so on. In addition, PSS is also of great
concern in atomic and molecular physics and has been discussed in some special
atomic and molecular
potentials~\cite{PhysScr2007CSJia,PRA2008FLZhang,Fewbody2013}.

Since the recognition of PSS in the nuclear spectrum, comprehensive efforts
have been made to explore its origin until Ginocchio pointed out that PSS is a
relativistic symmetry in the Dirac Hamiltonian, which is exactly conserved
when the scalar and vector potentials satisfying $\Sigma(r) \equiv
S(r)+V(r)=0$~\cite{PRL1997Ginocchio_78_436}. He also revealed that the
pseudo-orbital angular momentum $\tilde{l}$ is nothing but the orbital angular
momentum of the lower component of the Dirac wave
function~\cite{PRL1997Ginocchio_78_436}, and there are certain similarities in
the relativistic single-nucleon wave functions of the corresponding pseudospin
doublets~\cite{PRC1998Ginocchio_57_1167}. However, there is no bound state in
the PSS limit. Later, Meng \textit{et al}. pointed out a more general
condition of $d \Sigma(r)/d r=0$, which can be approximately satisfied in
exotic nuclei with highly diffuse
potentials~\cite{PRC1998JMeng_58_628,PRC1999JMeng_59_154} and the onset of the
pseudospin symmetry to a competition between the pseudo-centrifugal barrier
(PCB) and the pseudospin-orbit (PSO) potential. Afterwards, PSS in nuclear
spectra have been studied extensively, such as PSS in deformed
nuclei~\cite{PRC1998Lalazissis_58_R45,PRC1999Sugawara_58_R3065,PRC2000Sugawara_62_054307,PRC2002Sugawara_65_054313,
PRC2004Ginocchio_69_034303,EPJA2012YWSun_48_18,PRL2014JYGuo}, spin symmetry
(SS) in anti-nucleon
spectra~\cite{PRL2003SGZhou_91_262501,PRC2005Mishustin_71_035201,EPJA2006He_28_265,EPJA2010HZLiang_44_119,
PRC2010Lisboa_81_064324}, PSS and SS in
hypernuclei~\cite{CPL2009CYSong_26_122102,PRC2017Sun_96_044312,JPG2017Lu_44_125104},
a perturbative interpretation of SS and
PSS~\cite{PRC2002Alberto_65_034307,IJMPD2004Alberto_13_1447,PRC2010Lisboa_81_064324,PRC2011Liang_83_041301,CPC2011FQLi_35_825},
and PSS in supersymmetric quantum
mechanics~\cite{PRL2004Leviatan_92_202501,PRL2009Leviatan_103_042502,PRC2013Liang_87_014334,PRC2013Shen_88_024311}.

PSS in bound states is always broken according to the conservation condition.
In contrast, the resonant states, which can be obtained in the PSS limit and
in finite-depth potentials, provide us with a better platform for the studying
PSS. In addition, resonant states play essential roles in exotic nuclei, where
the neutron or the proton Fermi surface is very close to the continuum
threshold. Here valence nucleons can be easily scattered to single-particle
resonant states in the continuum due to pairing correlations, and the
couplings between the bound states and the continuum become very
important~\cite{PRL1996Meng_77_3963,PRC1996Dobaczewski_53_2809,PRL1997Poschl_79_3841,
PPNP2006JMeng_57_470}. Therefore, the study of PSS in resonant states has
attracted increasing attention in recent years. Until now, there are already
some investigations of the PSS in the single-particle resonant states. PSS and
SS in nucleon-nucleus and nucleon-nucleon scattering have been investigated in
Refs.~\cite{PRL1999Ginocchio_82_4599,PRC2000HLeeb_62_024602,PRC2002Ginocchio_65_054002,PRC2004HLeeb_69_054608}.
In 2004, Zhang \textit{et~al.} confirmed that the lower components of the
Dirac wave functions for the resonant PS doublets also have similarity
properties~\cite{CPL2004Zhang_21_632}. Guo \textit{et~al.} investigated the
dependence of pseudospin breaking for the resonant states on the shape of the
mean-field potential in a Woods-Saxon form~\cite{PRC2005JYGuo_72_054319,
PLB2020XXShi,PLB2022QLiu} as well as on the ratio of neutron and proton
numbers~\cite{PRC2006JYGuo_74_024320}. In 2012, great progress has been
achieved by Lu \textit{et~al.} in Ref.~\cite{PRL2012BNLu_109_072501}, where
they gave a rigorous justification of PSS in single-particle resonant states
and shown that PSS in single-particle resonant states is also exactly
conserved under the same conditions as PSS in bound states, i.e.,
$\Sigma(r)=0$ or $d\Sigma(r)/dr=0$~\cite{PRL2012BNLu_109_072501}. However, the
wave functions of the PS partners in the PSS limit are still absent. And also
their research is mainly based on a radial square-well
potential~\cite{PRC2013BNLv_88_024323}. Furthermore, a uniform description for
the conservation and breaking of PSS, i.e., from the PSS limit to cases with
finite-depth potentials is highly expected.

In this work, we will illustrate the exact conservation and breaking of PSS in
the nuclear single-particle states in spherical Woods-Saxon potentials. The
Green's function (GF) method~\cite{Book2006Eleftherios-GF,
PRB1992Tamura_45_3271,PRA2004Foulis_70_022706,PRC2011ZhangY_83_054301,Sci2016Sun_46_12006}
is employed. This method has been confirmed to be one of the most efficient
tools for studying the single-particle resonant states, because it has the
following advantages: the bound single-particle and the resonant states are
treated on the same footing, the energies and widths for all resonances are
precisely determined regardless of their widths, and the spatial density
distributions are properly
described~\cite{PRC2014TTSun_90_054321,PRC2020TTSun,CPC2020CChen,NST2021YTWang}.
Besides, this method can describe the resonant states in any potential without
any requirement on the potential shape.

This paper is organized as follows. The theoretical framework of the Green's
function method is briefly presented in Section \ref{sec:the}. Section
\ref{sec:num} is devoted to the discussion of the numerical results, where the
exact conservation and breaking of the PSS in single-particle resonant states
are illustrated by analyzing the energy and width splittings and the density
distributions. Finally, a summary is given in Section \ref{sec:con}.

\section{\label{sec:the}Theoretical framework}
In a relativistic framework, nucleons are Dirac spinors moving in a mean-field
potential with an attractive scalar potential $S({\bm r})$ and a repulsive
vector potential $V({\bm r})$~\cite{PPNP1996Ring_37_193}. The Dirac equation
for a nucleon reads
\begin{equation}
[\bm{\alpha}\cdot\bm{p}+V(\bm{r})+\beta(M+S(\bm{r}))]\psi_n(\bm{r})=\varepsilon_n\psi_n(\bm{r}),
\label{EQ:Dirac}
\end{equation}
where $\bm{\alpha}$ and $\beta$ are the Dirac matrices and $M$ is the nucleon
mass. Based on the Dirac Hamiltonian $\hat{h}(\bm{r})$, a relativistic
single-particle Green's function $\mathcal{G}(\bm{r},\bm{r'};\varepsilon)$ can
be constructed, which obeys
\begin{equation}
[\varepsilon-\hat{h}(\bm{r})]\mathcal{G}(\bm{r},\bm{r}';\varepsilon)=\delta(\bm{r}-\bm{r}').
\label{Eq:GF_define}
\end{equation}
With a complete set of eigenstates $\psi_{n}(\bm{r})$ and eigenvalues
$\varepsilon_{n}$, the Green's function can be simply represented as
\begin{equation}
\mathcal{G}(\bm{r},\bm{r}';\varepsilon)=\sum_n\frac{\psi_{n}(\bm{r})\psi_{n}^{\dag}(\bm{r}')}{\varepsilon-\varepsilon_{n}},
\label{EQ:GF}
\end{equation}
which is a $2\times2$ matrix because of the upper and lower components of the
Dirac spinor $\psi_n(\bm{r})$. Equation (\ref{EQ:GF}) is fully equivalent to
Eq.~(\ref{Eq:GF_define}).

For a spherical nucleus, the Green's function can be expanded as
\begin{equation}
\mathcal{G}({\bm r},{\bm r'};\varepsilon)=\sum_{\kappa m}Y_{jm}^{l}(\theta,\phi)\frac{\mathcal{G}_{\kappa}(r,r';\varepsilon)}{rr'}Y_{jm}^{l*}(\theta',\phi'),
\end{equation}
where $Y_{jm}^{l}(\theta,\phi)$ is the spin spherical harmonic,
$\mathcal{G}_{\kappa}(r,r';\varepsilon)$ is the radial Green's function, and
the quantum number $\kappa=(-1)^{j+l+1/2}(j+1/2)$. The
Eq.~(\ref{Eq:GF_define}) can be reduced as
\begin{equation}
\left[\varepsilon-\left(
                    \begin{array}{cc}
                      \Sigma(r) & -\frac{d}{dr}+\frac{\kappa}{r} \\
                      \frac{d}{dr}+\frac{\kappa}{r} & \Delta(r)-2M \\
                    \end{array}
                  \right)
\right]\mathcal{G}_{\kappa}(r,r';\varepsilon)
=\delta(r-r')I,
\label{EQ:rGF}
\end{equation}
where $\Sigma(r)\equiv V(r)+S(r)$, $\Delta(r)\equiv V(r)-S(r)$, and $I$ is a
two-dimensional unit matrix. A radial Green's function
$\mathcal{G}_{\kappa}(r,r';\varepsilon)$ could be constructed with exact
asymptotic behaviors of the Dirac wave functions for bound states and
continuum. For these details, please see
Refs.~\cite{PRC2014TTSun_90_054321,Sci2016Sun_46_12006}

To study the conservation and breaking of PSS in resonant states, radial
Woods-Saxon potentials are considered both for $\Sigma(r)$ and $\Delta(r)$,
\begin{equation}
\Sigma(r)=\frac{C}{1+e^{(r-R)/a}},~~\Delta(r)=\frac{D}{1+e^{(r-R)/a}}.
\label{EQ:WS}
\end{equation}
Here, the potential depths $C=-66$~MeV and $D=650~$MeV, the width $R=7$~fm,
and the diffusivity parameter $a=0.3$~fm are adopted.

\section{\label{sec:num}Results and discussions}

\begin{figure}[ht!]
\includegraphics[width=0.45\textwidth]{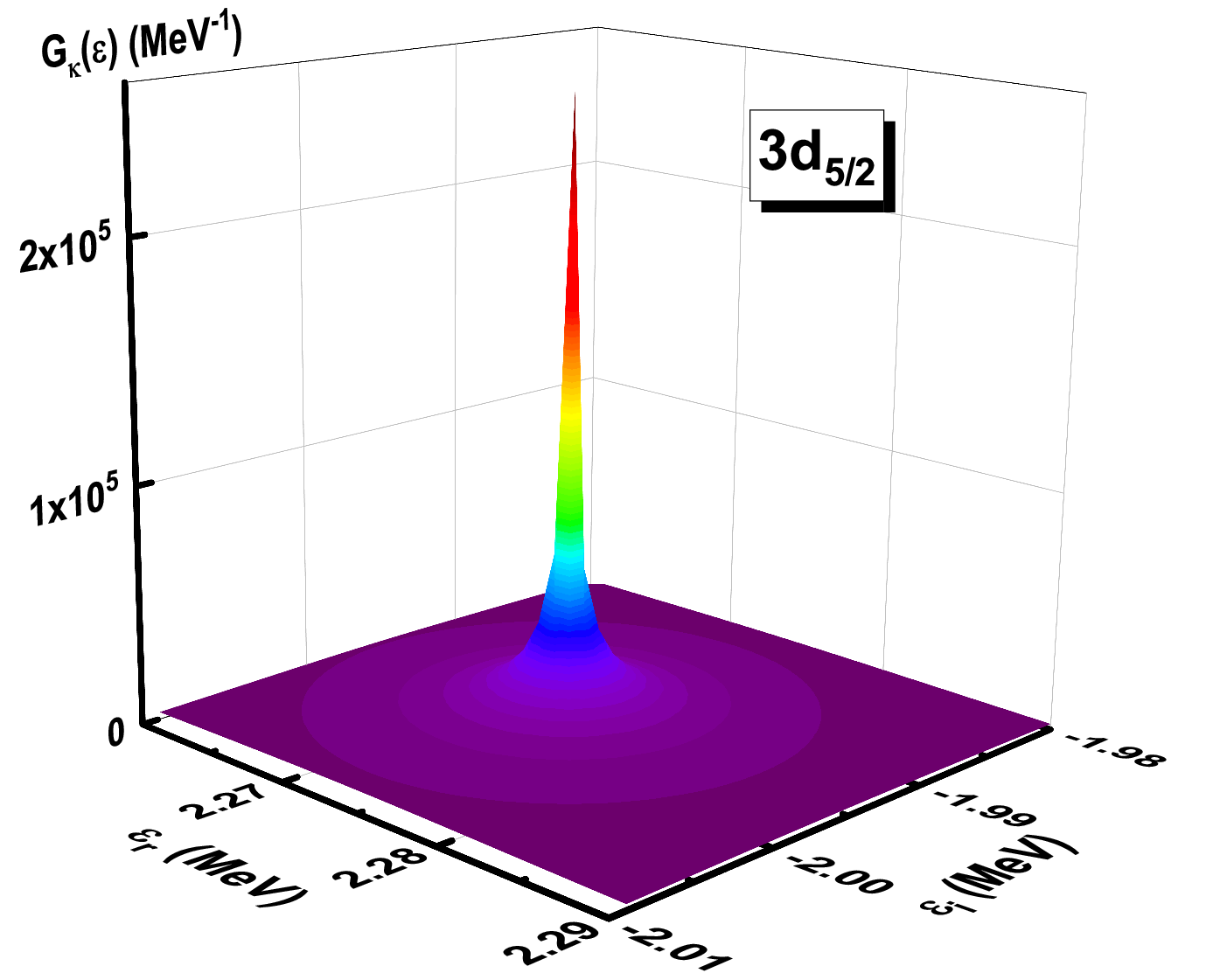}
\caption{(Color online) The single-particle resonant state $3d_{5/2}$ located in
the fourth quadrant of the complex energy plane
$\varepsilon=\varepsilon_r+i\varepsilon_i$ determined by searching for the
poles of the Green's function $G_{\kappa}(\varepsilon)$.}
\label{res}
\end{figure}

On the single-particle complex energy plane, bound and resonant states are
distributed on the negative real energy axis and in the fourth quadrant,
respectively. The energy $\varepsilon_n$ is real for bound states while
complex for resonant states and in the latter case $\varepsilon_n=E-i\Gamma/2$
with $E$ and $\Gamma$  being the resonant energy and the width respectively.
As shown in Eq.~(\ref{EQ:GF}) these eigenvalues are the poles of the Green's
function. Thus, in Refs.~\cite{PRC2020TTSun,CPC2020CChen,NST2021YTWang} it has
been proposed to determine the single-particle energies $\varepsilon_{n}$ by
searching for the poles of the Green's function. In practice, one can do this
by calculating the integral function of the Green's function
$G_{\kappa}(\varepsilon)$ for each partial wave $\kappa$ at different energies
$\varepsilon$~\cite{NST2021YTWang}
\begin{equation}
G_{\kappa}(\varepsilon)=\int dr \left( |\mathcal{G}_{\kappa}^{(11)}(r,r;\varepsilon)|+|\mathcal{G}_{\kappa}^{(22)}(r,r;\varepsilon)|\right),
\end{equation}
where $|\mathcal{G}_{\kappa}^{(11)}(r,r;\varepsilon)|$ and
$|\mathcal{G}_{\kappa}^{(22)}(r,r;\varepsilon)|$ are the moduli of the Green's
functions respectively for the ``11'' and ``22'' matrix elements. To search
for the bound and resonant states, Green's functions in a wide energy range
are calculated by scanning the single-particle energy $\varepsilon$. For the
bound states, the energies $\varepsilon$ are taken along the negative real
energy axis. For the resonant states, the energies $\varepsilon$ are complex
$\varepsilon=\varepsilon_r+i\varepsilon_i$ which are scanned in the fourth
quadrant of the complex energy plane $\varepsilon$, both along the real and
imaginary energy axes. In Fig.~\ref{res}, the resonant parameters of the state
$3d_{5/2}$ are exactly determined to be $E=2.2728~$MeV and
$\Gamma/2=1.9949~$MeV by searching for the poles of the Green's functions in
the fourth quadrant of the complex energy plane $\varepsilon$, where a sharp
peak is observed at $\varepsilon_{r}=2.2728$~MeV and
$\varepsilon_{i}=-1.9949~$MeV. Calculations are done with an energy step of
$0.1$~keV for the integral functions $G_{\kappa}(\varepsilon)$ in a coordinate
space with size $R_{\mathrm{ max}}=20~$fm and a step of $dr=0.05$~fm. This
approach has been certified to be highly effective for all resonant states
regardless of whether they are wide or
narrow~\cite{CPC2020CChen,NST2021YTWang}.

\begin{figure}[h!]
\includegraphics[width=0.45\textwidth]{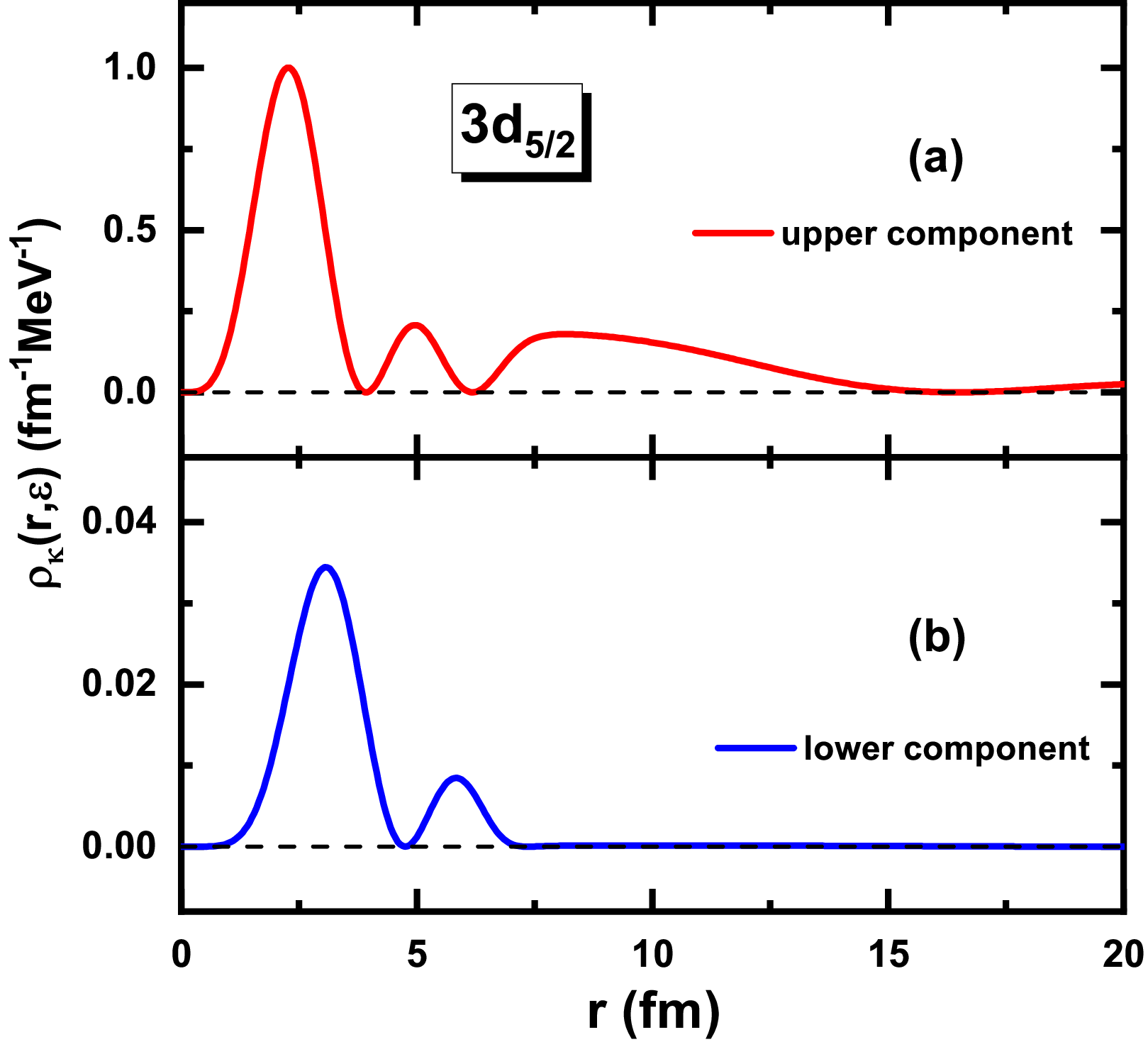}
\caption{(Color online) Density distributions $\rho_{\kappa}(r,\varepsilon)$
of the resonant state $3d_{5/2}$ with the contributions from the upper and
lower components of the Dirac wave functions.} \label{wf1}
\end{figure}

Besides, with the Green's function method, the density distributions in the
coordinate space can also be examined by exploring
$\rho_{\kappa}(r,\varepsilon)$ defined at the energy $\varepsilon=E$,
\begin{eqnarray}
&&\rho_{\kappa}(r,\varepsilon)\\
&=&-\frac{1}{4\pi r^2}\frac{1}{\pi}{\mathrm{Im}}\left[\mathcal{G}_{\kappa}^{(11)}(r,r;E)+\mathcal{G}_{\kappa}^{(22)}(r,r;E)\right],\nonumber
\end{eqnarray}
where the terms $\mathcal{G}_{\kappa}^{(11)}(r,r;E)$ and
$\mathcal{G}_{\kappa}^{(22)}(r,r;E)$ are respectively related to the upper and
lower components of the Dirac wave functions (c.f.~Eq.~(\ref{EQ:GF})). In
Fig.~\ref{wf1}, the density distributions for the resonant state $3d_{5/2}$
are also plotted, with the red and blue lines the contributions from the upper
and lower components. To better display the density distribution, here and
hereafter, we adjust the highest peak of $\rho_{\kappa}(r,\varepsilon)$ to be
$1.0$~fm$^{-1}\cdot$MeV$^{-1}$ and ensure relative sizes of the different
components remain unchanged. Since $3d_{5/2}$ is a low-lying resonant state,
the density distribution of the lower component is reduced to zero very soon
while a very slight oscillation can be observed in the upper component.

\begin{figure}[t!]
\includegraphics[width=0.45\textwidth]{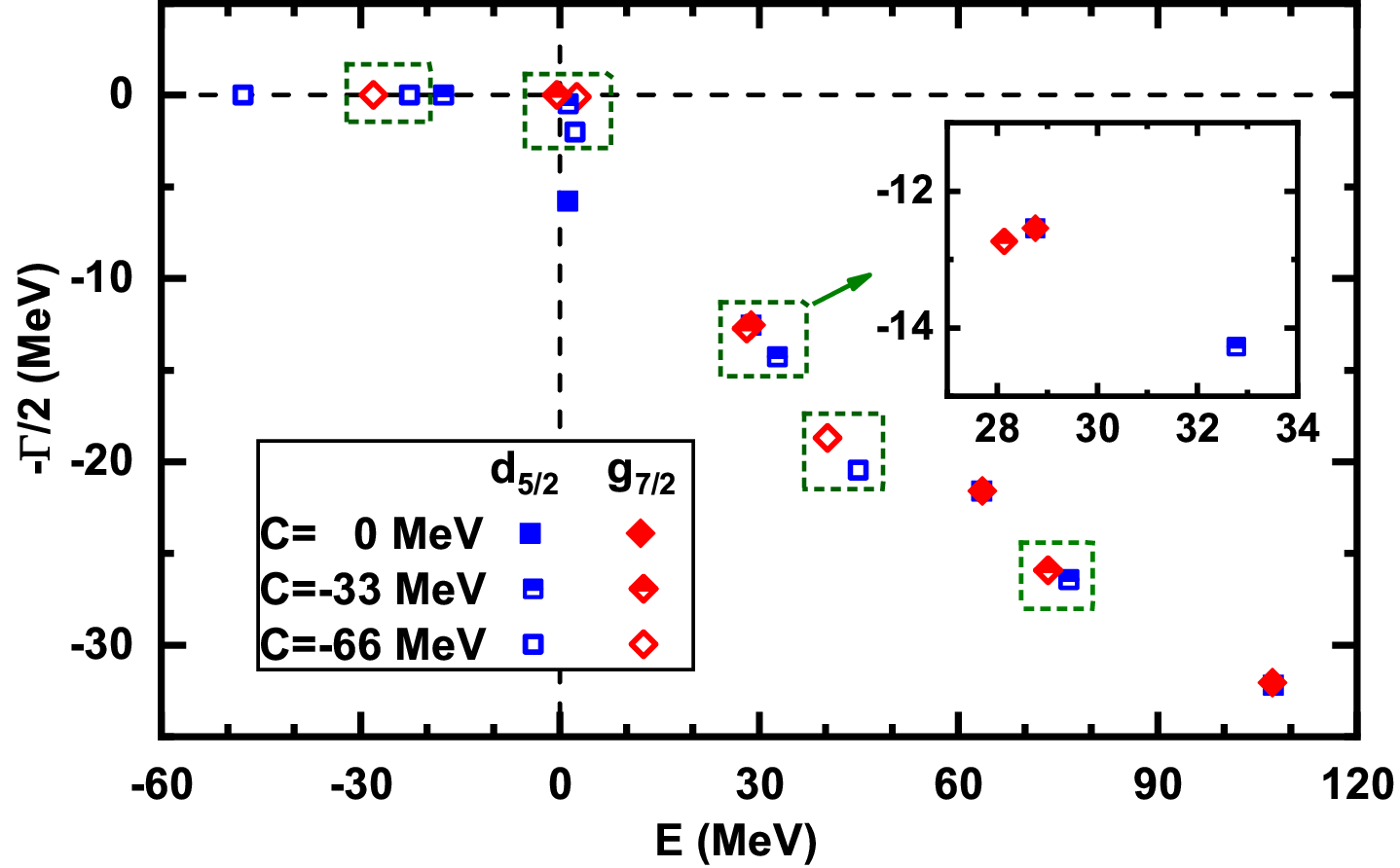}
\caption{(Color online) The poles of Green's functions on the complex energy plane in the Woods-Saxon potentials with
depth $C=0$~(solid symbols), $C=-33~$MeV~(half-filled symbols) and
$C=-66~$MeV~(empty symbols) for the PS partners $d_{5/2}$~(square) and
$g_{7/2}$~(diamond).}
\label{EWPSS}
\end{figure}

In the following, PSS in resonant states will be studied with the Green's
function method. In Fig.~\ref{EWPSS}, we show the solutions in different
potential depths on the complex energy plane for the PS doublets with
pseudospin angular momentum $\tilde{l}=3$, i.e., $d_{5/2}$ with $\kappa=-3$
and $g_{7/2}$ with $\kappa=4$. In the PSS limit, i.e., for the potential depth
$C=0$, all the roots are located in the lower half-plane, and there are no
bound states. Three pairs of resonant PS doublets with exactly the same energy
and width are obtained, indicating the exact conservation of PSS in resonant
states. Besides, one single intruder state $1d_{5/2}$ appears near the
continuum threshold. With finite potential depths, one finds the breaking of
the PSS with obvious energy and width splitting between the PS partners. More
in details, for most PS partners, $g_{7/2}$ with pseudospin $\tilde{s}=+1/2$
has lower energy and smaller width compared with the PS partner $d_{5/2}$ with
$\tilde{s}=-1/2$ due to the higher PCB potential of $g_{7/2}$. One exception
are the PS partners $(3d_{5/2}, 2g_{7/2})$ obtained for $C=-66$~MeV with the
energies $\varepsilon(3d_{5/2})=2.2728-i1.9949$~MeV and
$\varepsilon(2g_{7/2})=2.5422-i0.1019~$MeV, respectively. Meanwhile, PS
partners move down and some resonant PS partners evolve to be bound states.
For PS doublets with other values of $\tilde{l}$, similar behaviors concerning
the exact conservation and the breaking of the PSS could be observed.

\begin{figure}[t!]
\includegraphics[width=0.45\textwidth]{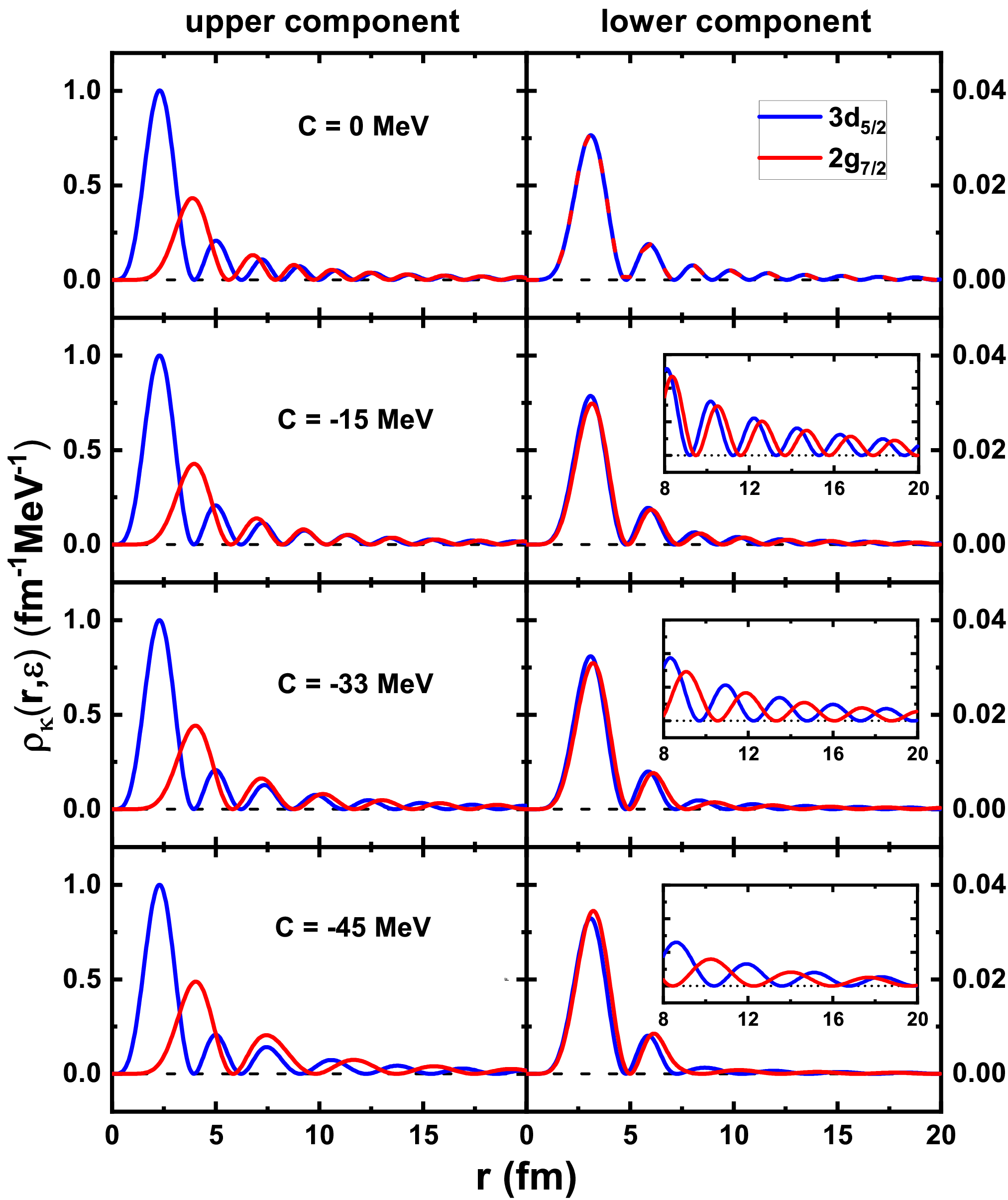}
\caption{(Color online). The density distributions $\rho_{\kappa}(r,\varepsilon)$ for the
PS doublets $3d_{5/2}$ and $2g_{7/2}$ for various depths of the potentials, from the PSS limit with $C=0~$MeV to the cases with $C=-15, -33, -45$~MeV. The densities plotted in the left
and right columns correspond to the upper and lower components of the
Dirac wave functions, respectively.}
\label{wf2}
\end{figure}

To study the conservation and breaking of PSS, the similarities of the lower
component of the Dirac wave functions for the PS doublets are also examined.
In Fig.~\ref{wf2}, the spacial density distributions
$\rho_{\kappa}(r,\varepsilon)$ of the PS partners $3d_{5/2}$ and $2g_{7/2}$
are plotted for different values of the potential depth $C$ . The left and
right columns present the contributions from the upper and lower components of
the Dirac wave functions. In the PSS limit, i.e., for $C=0$, the density
distributions for the PS partner are identical for the lower component while
they differ by one node for the upper component. This provides direct evidence
for the exact conservation of PSS in resonant states and also certifies that
PSS is a symmetry of the Dirac Hamiltonian related to the lower component of
the Dirac spinor. For potentials with finite depth, the lower components of
the density distributions of the PS partners are no longer identical, but they
still show a strong similarity. Their difference is manifested as an obvious
phase shift, which increases with the growing of the potential depths. For
example, when the potential depth $C=-45~$MeV, we observe a phase shift of
almost one $\pi$ between the PS partners for the density distributions outside
the potential, in the area of $r>8.0~$fm. Possibly, this phase shift may be
extracted and confirmed by low-energy neutron-nuclei scattering experiments.

\section{\label{sec:con} Summary}

In summary, the conservation and breaking of PSS in nuclear single-particle
states are investigated within a relativistic framework by exploring the poles
of the Green's function in spherical Woods-Saxon potentials of different
depths. The Green's function method allows a precise determination of the
energies and the widths for all the resonances and a proper description of the
spatial density distributions. Therefore, it provides an excellent platform
for studying the breaking and the restoration of PSS. In the PSS limit, i.e.,
for $\Sigma(r)\equiv V(r)+S(r)=0$, the PSS in resonant states is confirmed to
be strictly conserved with exactly the same energy and width for the PS
partners. Besides, we also find identical density distributions of the lower
components for the first time. This provides direct evidence that the PSS is a
relativistic dynamical symmetry connected with the lower component of the
Dirac spinor. For potentials with finite depth, PSS is broken, combined with
an apparent splitting of the energy and the width for the PS partners and a
phase shift between the spatial density distributions of the lower components.

\section*{ACKNOWLEDGMENTS}
This work was partly supported by the National Natural Science Foundation of
China (No.~U2032141 and No.~12375126), the Open Project of Guangxi Key
Laboratory of Nuclear Physics and Nuclear Technology (No. NLK2022-02), the
Central Government Guidance Funds for Local Scientific and Technological
Development, China (No. Guike ZY22096024), the Fundamental Research Funds for
the Central Universities, and by the Deutsche Forschungsgemeinschaft (DFG,
German Research Foundation) under Germanys Excellence Strategy
EXC-2094-390783311, ORIGINS.

\balance

\end{document}